\documentclass[11pt,twoside]{article}
\usepackage{./asp2014}

\aspSuppressVolSlug
\resetcounters

\begin{document}

\title{\textbf{\textit{Gaia}}, Stellar Populations and the Distance Scale}
\author{Gisella~Clementini,$^1$ Alessia~Garofalo,$^{1,2}$ Tatiana~Muraveva,$^1$ and Vincenzo Ripepi,$^3$
\affil{$^1$INAF- Osservatorio di Astrofisica e Scienza dello Spazio, Bologna, Italy; \email{gisella.clementini@oabo.inaf.it}\\
\email{tatiana.muraveva@oabo.inaf.it}}
\affil{$^2$Dipartimento di Fisica e Astronomia-Universit\`a di Bologna, Bologna, Italy; \email{alessia.garofalo@unibo.it}}
\affil{$^3$INAF-Osservatorio Astronomico di Capodimonte, Naples,  Italy; \email{ripepi@na.astro.it}}}

\paperauthor{Gisella~Clementini}{gisella.clementini@oabo.inaf.it}{0000-0001-9206-9723}{INAF}{Osservatorio di Astrofisica e Scienza dello Spazio}{Bologna}{Bo}{40129}{Italy}
\paperauthor{Alessia~Garofalo}{alessia.garofalo@unibo.it}{ORCID_Or_Blank}{Universit\'a di Bologna}{Dipartimento di Fisica e Astronomia}{Bologna}{Bo}{40129}{Italy}
\paperauthor{Tatiana~Muraveva}{tatiana.muraveva@oabo.inaf.it}{ORCID_Or_Blank}{INAF}{Osservatorio di Astrofisica e Scienza dello Spazio}{Bologna}{Bo}{40129}{Italy}
\paperauthor{Vincenzo~Ripepi}{ripepi@na.astro.it}{ORCID_Or_Blank}{INAF}{Osservatorio Astronomico di Capodimonte}{Naples}{Na}{80131}{Italy}

\begin{abstract}
We discuss the impact that \textit{Gaia}, a European Space Agency (ESA) cornerstone mission that has been in scientific operations since July 2014, is expected to have on the definition of the cosmic distance ladder and the study of resolved stellar populations in and beyond the Milky Way, specifically focusing on results based on Cepheids and RR Lyrae stars.

\textit{Gaia} is observing about 1.7 billion sources, measuring their position, trigonometric parallax, proper motions and time-series photometry in 3 pass-bands down to a faint magnitude limit of $G \sim$21 mag. Among them are thousands of Cepheids and hundreds of thousands of RR Lyrae stars. After a five years of mission operations the parallax errors are expected to be of about 10 microarcsec for sources brighter than $V \sim$ 12, 13 mag.
This will allow an accurate re-calibration of the fundamental relations that make RR Lyrae stars and Cepheids primary standard candles of the cosmic distance ladder and will provide a fresh view of the systems and structures that host these classical pulsators.
Results for Cepheids and RR Lyrae stars published in \textit{Gaia} Data Release 1 (DR1) are reviewed along with some perspectives on \textit{Gaia} DR2, scheduled for 25 April 2018, which will contain parallaxes based only on \textit{Gaia} measurements and a first mapping of full-sky RR Lyrae stars and Cepheids.

\end{abstract}

\section{Introduction}

The \textit{Gaia} mission has been surveying the whole sky since July 2014, measuring astrometric parameters (positions, parallaxes and proper motions) and collecting multi-epoch photometry in three different pass-bands (\textit{Gaia} $G$, $G_{BP}$ and $G_{RP}$ ) of sources within a limiting magnitude of $G$ = 21 mag and a bright limit of $G \approx$ 3 mag 
that transit across its field of view.   The spacecraft  also simultaneously collects  spectroscopy with the Radial Velocity Spectrometer (RVS) of the sources brighter that $V \sim$ 16 mag. At the end of a five year nominal duration of the mission each source will have on average 70 photometric measurements (over 200 observations for sources at $|$DEC$|$=45 $\pm$ 10 deg)  and about 40 spectroscopic measurements with the RVS.
This makes \textit{Gaia} a most powerful tool  to discover and characterise variable sources. About  1.7 billion sources are being measured by {\it Gaia}, among which are thousands of Cepheids and hundred of thousands RR Lyrae stars in the Milky Way (MW) and its nearest neighbours.
  
Astrometric, photometric and spectroscopic performances of \textit{Gaia} vary depending on the source magnitude and spectral type. Estimated end-of mission (e.o.m.) values are summarised on the ESA  \textit{Gaia} web page  (https://www.cosmos.esa.int/web/gaia/ science-performance). In particular, for  the typical spectral types of RR Lyrae stars and Cepheids  e.o.m. standard errors of the integrated photometry  are expected to be of  0.2 millimag in $G$  and 1 millimag in $G_{BP}$ and $G_{RP}$  at $G\sim$ 15 mag,  and 3.7, 56, and 48 millimag, respectively, at $G\sim$ 20 mag. Astrometric standard errors are of 24 $\mu$as at $V\sim 15$ mag and 540 $\mu$as at $V\sim 20$ mag and radial velocity errors are below 1 km s$^{-1}$  for $V <$ 12 mag and below 15 km s$^{-1}$  for $V <$ 15 mag.

A major asset of \textit{Gaia} observations  is that there are no proprietary data rights.  \textit{Gaia}  data become public as soon as they have been fully processed and properly validated. 
Release of the  \textit{Gaia} final catalogue of data collected over the 5 year nominal duration of the mission is scheduled for 2022. However,  publication of preliminary data products is also anticipated through a number of intermediate data releases,  the first  
of which (\textit{Gaia} Data Release 1 - DR1) took place on 14  September 2016,  a second release (\textit{Gaia} DR2) will occur on 25 April 2018 and a third one 
(\textit{Gaia} DR3) is currently foreseen in the second half of 2020. 

\section{Cepheids and RR Lyrae stars with Gaia}

The potential of  \textit{Gaia} in the field of variable stars such as Cepheids and RR Lyrae stars is enormous, encompassing the complete census of such variables in and beyond the MW and the direct measure through parallax of their distances, hence  absolute magnitude, locally and farther than the  Magellanic Clouds, although with relatively larger individual errors in the latter case.
According to current  estimates,  \textit{Gaia} is expected to increase up to a factor of ten the number of known Classical Cepheids in the MW and to enlarge the number of known Galactic RR Lyrae stars well beyond the current value of more than a hundred thousands.   

RR Lyrae stars and Cepheids are the most important stellar distance indicators in the Local Group (LG) and beyond (e.g. \citealt{muraveva2015}, and references therein). They also are fundamental tracers of old ($t >$10 Gyr, RR Lyrae) and young (50 $< t <$ 500 Myr, Classical Cepheids) stellar populations, complementing the color-magnitude diagram (CMD) in recovering the star formation histories (SFH) of the resolved stellar populations in galaxies. 
The complete census of Galactic Cepheids and RR Lyrae stars along with an unprecedented precision and accuracy of {\it Gaia} parallax measurements for local Cepheids and RR Lyrae 
will allow a breakthrough in our understanding of the MW structure, dynamics and formation, and a global re-assessment of the whole cosmic distance ladder from local to cosmological distances. 
Concurrently, RR Lyrae stars and Cepheids can be used to test  the accuracy and precision of the parallaxes  and to assess systematics, biases  
and correlations in the {\it Gaia}  parallax data.

By observing thousands of Galactic and Magellanic Cloud pulsators,  
{\it Gaia} will allow us to calibrate with unprecedented precision and statistics the fundamental relations connecting the pulsation period $P$ to the mean magnitudes and colors of Cepheids and RR Lyrae  stars. This specifically implies using  {\it Gaia} parallaxes to calibrate  
the optical and near/mid infrared Period-Luminosity ($PL$), Period-Luminosity-Color ($PLC$) and  Period-Wesenheit ($PW$) relations of Classical and Type~II Cepheids; the optical Luminosity-Metallicity relation ($M_{V}$ - [Fe/H]) and the near/mid-infrared $PL$ and Period-Luminosity-Metallicity ($PLZ$) relations of RR Lyrae  stars; to  study in detail the systematics affecting these relations and,   to test their universality across different stellar systems. It also opens the way to possibly extend those relations to  the "ultra-long" period ($ULP$) Cepheids (Cepheids with periods longer than  80 days and about 2-4 mag brighter than Classical Cepheids) that could serve as stellar standard candles reaching cosmologically relevant distances  $D \geq 100 $ Mpc, in only one step. 
Furthermore,  the parallax information combined with chemistry from the RVS  
and complementary  spectroscopy from ongoing and future ground-based surveys  will allow us to study the metallicity dependence of the $PL$/$PW$ relations and at the same time  determine the metallicity distribution of young (with Classical Cepheids) and old (with RR Lyrae stars) stellar populations in the host systems. 
{\it Gaia} will also make it possible to directly calibrate via parallax the Tip of the Red Giant Branch (TRGB) luminosity for a number of close by stellar systems.
 Secondary indicators 
peering into the  unperturbed Hubble Flow such as the Type Ia supernovae (SNe Ia)  and the surface brightness fluctuations (SBF) will be re-calibrated on the {\it Gaia}-based relations of Cepheids and RR Lyrae stars, 
 thus setting the basis for a measurement of the Hubble constant with 1\% accuracy and at the same cementing \textit{Gaia}'s impact in the JWST, LSST and E-ELT era.

On the other hand, {\it Gaia}'s collecting time-series photometry over the whole sky will allow the discovery and characterisation of thousands of new RR Lyrae stars and Cepheids in still unexplored regions of the MW and its satellites. By combining positions and proper motions  of  RR Lyrae stars and Cepheids observed  by \textit{Gaia} down to $G\sim$ 21 mag it will be possible to trace the corresponding parent stellar populations, map the  3-dimensional geometry  the MW and its neighbours,    
specifically tracing the distribution of young 
and old  stars, even in crowded regions,  and 
then comparing them with existing theoretical models in order to understand their formation and evolution. For instance, this study will be particularly important to better understand the  Magellanic System (Large Magellanic Cloud -- LMC, Small Magellanic Cloud -- SMC, and the bridge connecting them) and its interaction with the MW.

With the {\it Gaia} DR2 
release of a first mapping of full-sky RR Lyrae stars it will already be possible to use them to map  
the Galactic halo, the disk and the bulge,  to  unveil new streams, possibly discover new satellites, to study radial trends, extra-tidal stellar populations around globular clusters,  and to trace tidal streams and   halos around satellite galaxies
within the reach of {\it Gaia}.  The study of the specific properties of these variables (period, metallicity, etc.) will  help identifying  the ``building-blocks'' that have contributed to building up the MW halo, whether ``classical'' satellites  like Sagittarius, Carina, Sculptor, Ursa Minor and the two Magellanic Clouds, or  members of the ultra-faint class of dwarfs recently discovered around the MW,  ultimately providing hints about how our Galaxy has been assembled.  
For years,  the classical CMD-reconstruction method via the $\chi^{2}$ minimization 
has been used to derive the space-resolved star formation histories (SFH) of nearby galaxies, 
setting the foundations of the so-called "near-field cosmology."  An example of how powerful and innovative the approach of combining  
the SF recovery based on  the CMD-reconstruction  and the analysis of the variable star population can be is, for instance, the study of the ultra-faint dwarf galaxy Leo~T  by \citet{clementini2012}.  
The precise photometry and distances for thousands of variables and several million constant stars provided by {\it Gaia} will allow extending this combined approach  to a large fraction of the MW, and to improve the line-of-sight resolution in the analysis of the nearest galaxies.

Finally, \textit{Gaia}'s astrometry, spectroscopy,  time-series photometry and radial velocity measurements covering the entire sky will also enable us to constrain stellar intrinsic parameters, such as  the  mass, luminosity and effective temperature, 
 to test and improve the input physics of evolutionary and pulsation models, as well as to improve the physics of indirect techniques used to measure distances beyond the reach of  \textit{Gaia} parallaxes.

\section{Cepheids and RR Lyrae stars in  {\textbf{\textit{Gaia}}} Data Release 1}

\textit{Gaia}  Data Release 1 (DR1) published positions and $G$ magnitudes for about 1 billion stars from observations taken in 14 months from July 2014 and September 2015  (\citealt{gai16a}, 2016b)
and five-parameter astrometry for about 2-million sources in common between the Thyco-2 and  \textit{Gaia}  catalogues, obtained as a joint astrometric solution of the 
Thyco and \textit{Gaia} measurements (TGAS), specific for this first release. The TGAS sample includes parallaxes for more than 700 Galactic Cepheids and RR Lyrae stars. 

\articlefigure[trim=50 130 50 130, width=0.6\textwidth]{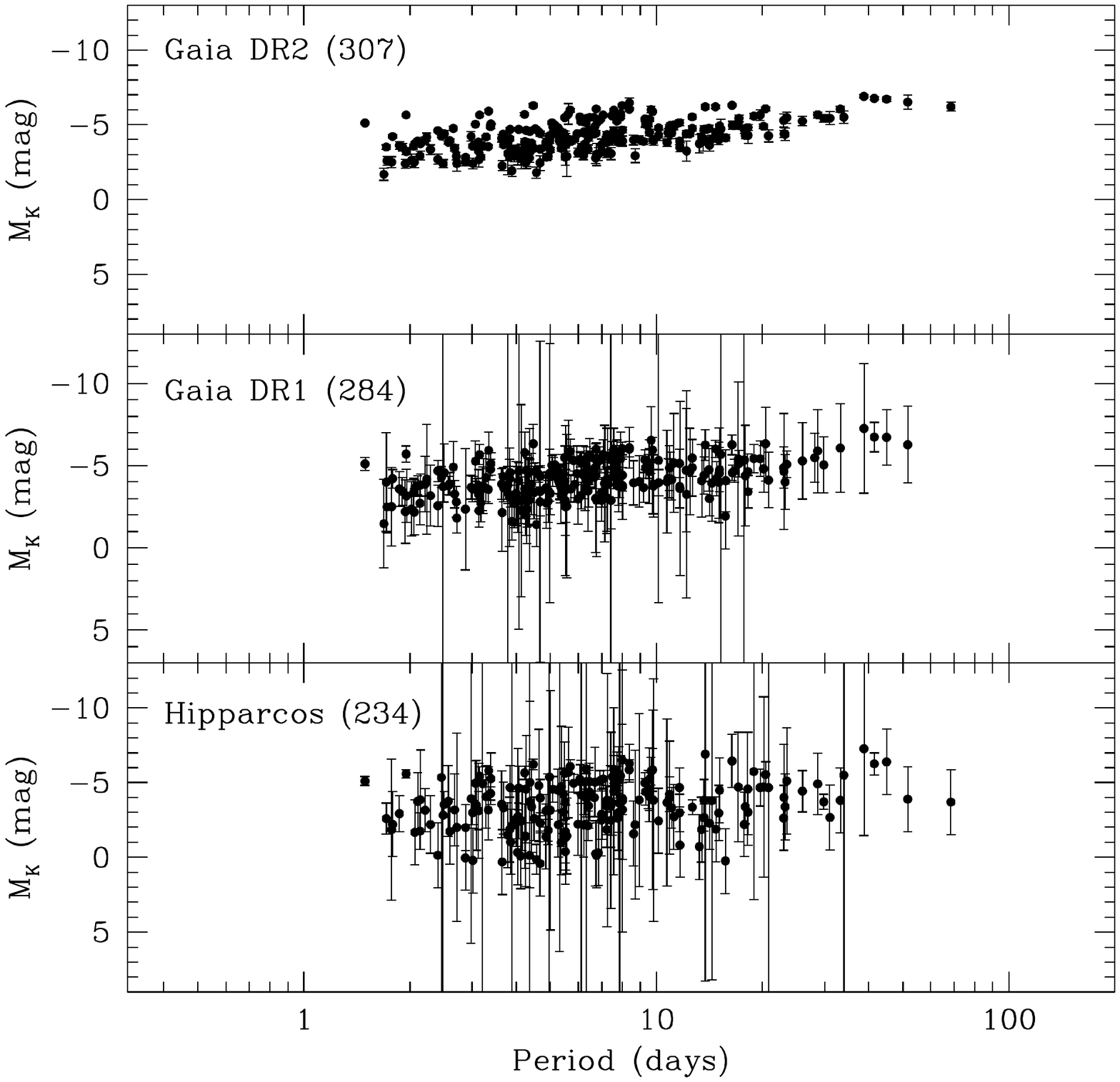}{fig1}{ Classical Cepheid $PL$  relation in the $Ks$-band using Hipparcos (bottom panel) and TGAS (middle panel) 
parallaxes (adapted from Fig. 20 in \citealt{gai17}). The top panel shows 
the Cepheid $PL$ relation based on  the {\it Gaia}-only parallaxes released in DR2 (Ripepi et al., in prep).}
\articlefigure[trim=50 130 50 130, width=0.6\textwidth]{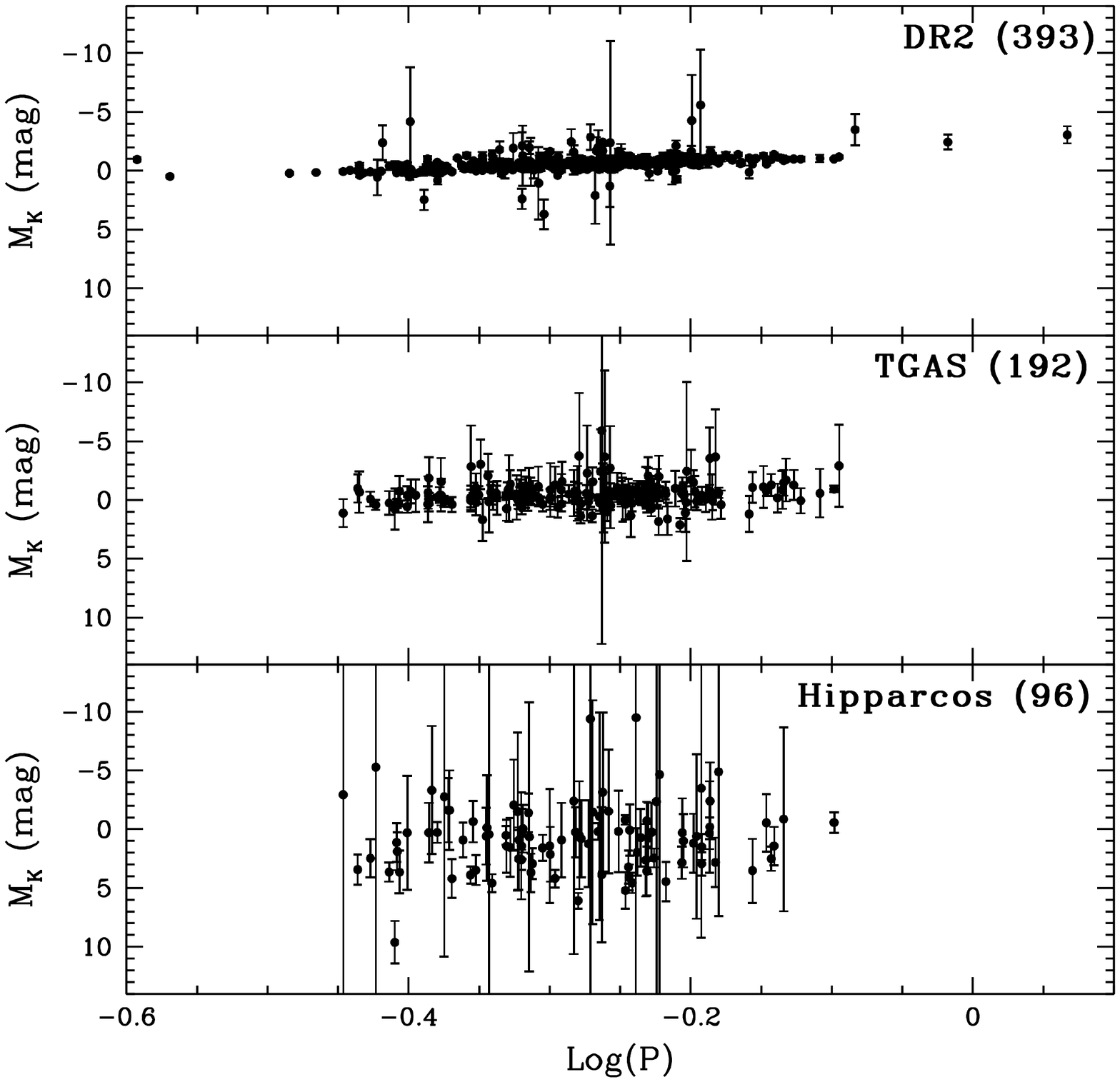}{fig2}{Same as in Fig.~\ref{fig1} but for RR Lyrae stars 
(adapted from Fig. 23 in \citealt{gai17}). The top panel shows the RR Lyrae $PL$ relation based on  the {\it Gaia} DR2 parallaxes (adapted from \citealt{muraveva2018}).
}

A  taste of {\it Gaia}'s potential relative to Cepheid and RR Lyrae calibrations has been shown by \citet{gai17} who used  the TGAS parallaxes 
along with literature ($V, I, J, {K_\mathrm{s}}, W_1$) photometry and spectroscopy, to calibrate the zero point of  
the $PL$ and $PW$ relations of Classical and Type~II Cepheids, and the near/mid-infrared $PL$, $PLZ$ and optical $M_{V}$ - [Fe/H] relations of RR Lyrae stars. 
Figure~\ref{fig1} shows a comparison of  the  $Ks$-band  $PL$ relations defined by Galactic Classical Cepheids in DR1 when using  the Hipparcos (bottom panel) and TGAS (middle panel) parallaxes, respectively. Figure~\ref{fig2} shows the same comparison for RR Lyrae stars.  Although this  is not comparable to the final {\it Gaia} precision, 
it already represents a significant general improvement with respect to Hipparcos parallaxes, particularly for RR Lyrae stars.
A further step forward is represented by {\it Gaia}-only astrometry released in DR2 (see upper panels of Figs.~\ref{fig1}  and ~\ref{fig2}).

 The fit the $PL$ relations, the \citet{gai17} applied  different techniques that operate either in distance/absolute magnitude or  in parallax space directly. 
However, we note that the direct transformation to absolute magnitude by parallax inversion is not possible if parallaxes are negative, thus biasing the sample 
due to the selective removal of distant sources with close to zero/negative parallaxes. Additionally, due to the presence of the 
logarithmic term in the relation between parallax and absolute magnitude, symmetrical errors in the parallaxes 
become asymmetric errors in magnitude. Therefore, working in parallax space, and adopting for instance a Bayesian approach, 
is definitely preferred, as it allows one to maintain symmetrical the errors and use negative parallax values. 
The hierarchical Bayesian model used in  \citet{gai17} is fully described by \citet{delgado2018}.

The publication of  variability data products was originally scheduled  to start much later than the {\it Gaia} first  data release; however,  in  
advance of schedule,  in DR1  
 $G$-band time series photometry  and characteristic parameters  were released for  a small sample of 3194 LMC RR Lyrae stars and Cepheids  
that \textit{Gaia}  observed at high cadence during the first 28 days of scientific operation in  Ecliptic Poles Scanning  Law.
A detailed description of the specific processing and main characteristics of  the RR Lyrae stars and Cepheids released in  
DR1 can be found in \citet{clementini2016}, whereas the general approach of variability analysis and classification of the \textit{Gaia} data is presented  in \citet{eyer2017}. 
Nice examples of the {\it Gaia} photometric quality at the faint magnitudes  of the RR Lyrae stars in the LMC are shown in Fig. 23 of \citet{clementini2016},  
whereas Fig.~32 in that paper showcases the potential of {\it Gaia} in relation to RR Lyrae stars as tracers of galaxy halos.  

\section{Cepheids and RR Lyrae stars in {\textbf{\textit{Gaia}}} Data Release 2}

Data Release 2 (DR2) will contain astrometry, photometry, and radial velocities  from the processing of all-sky sources observed over 22 months between July 2014 and May 2016. 
The content of the release is described in some detail in the ESA {\it Gaia} web page (https://www.cosmos.esa.int/web/gaia/dr2).  Among other products,  DR2 will include the five-parameter astrometric solution (positions, parallaxes and proper motions) based only on {\it Gaia} measurements for more than 1.3 billion sources with $G$ magnitudes from about 3 to  21 mag, median radial velocities  for more than 7.2 million stars with G magnitudes between about 4 and 13 mag, $G$ magnitudes for about 1.7 billion sources and $G_{BP}$, $G_{RP}$ photometry for about  1.4 billion sources. With specific relevance to the discussion presented in this paper,  {\it Gaia} DR2 will also publish the classification (and multi-band time-series photometry) for more than half a million variable sources of which a good fraction are RR Lyrae stars spread over  the whole celestial sphere. The line-of-sight extinction A$_G$ and the metallicity [Fe/H] inferred from the pulsation characteristics (period, amplitude and Fourier parameters of the light curve) will also be released for a significant number of these RR Lyrae  stars (along with A$_G$ values for some  Classical Cepheids). They will probe the dust and  metallicity distribution of the old stellar component in the MW.  

Uncertainty of the DR2 parallaxes is around 0.04 mas for sources  brighter the $G$ = 15 mag, and on the order of 0.7 mas at $G$ = 20 mag.
\articlefigure[trim=0 0 0 -10, width=0.9\textwidth]{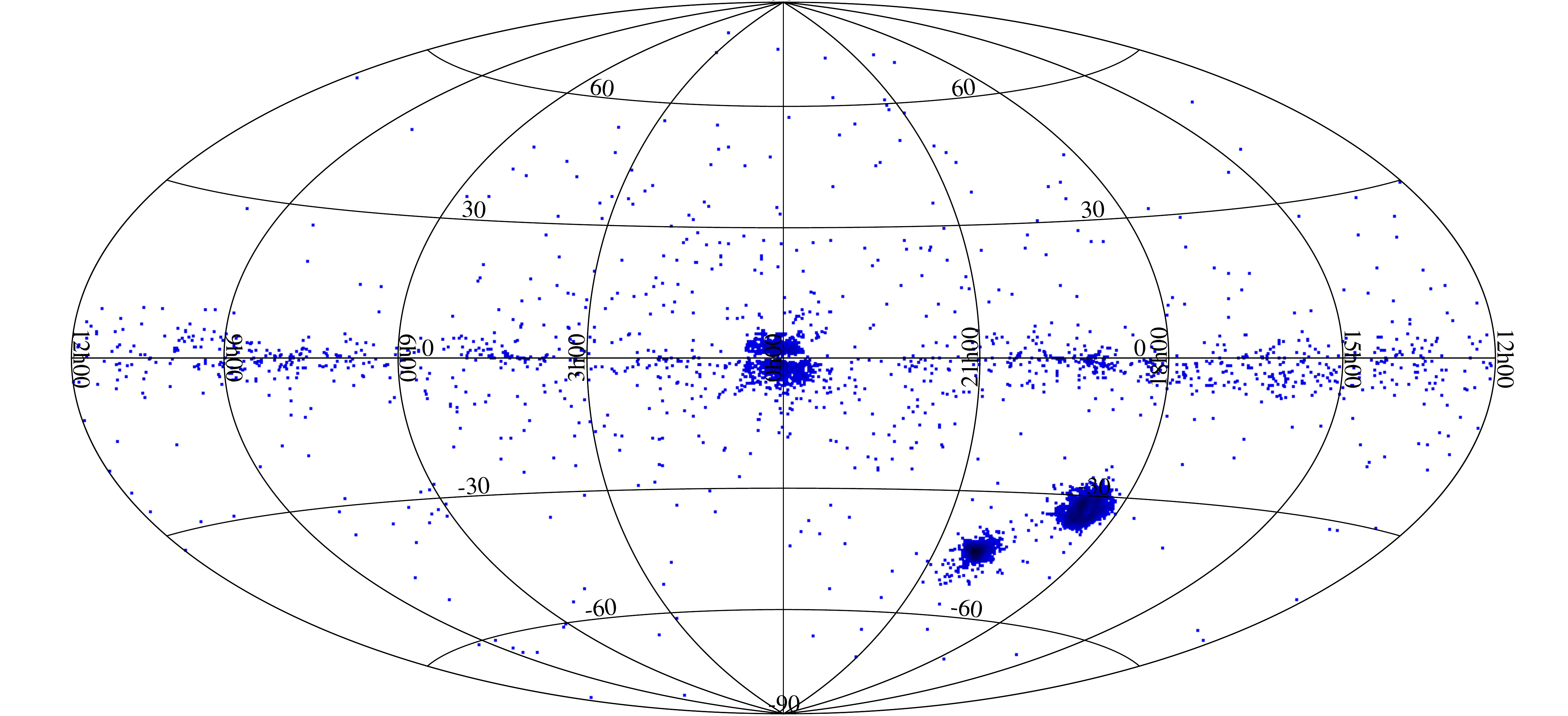}{fig3}{Sky map, in Galactic coordinates,  of literature Cepheids. See Fig.~39 of  \citet{clementini2018}
for an updated version of this figure showing  more than  350  new Cepheids discovered by {\it Gaia} and released in DR2.
}
With these first {\it Gaia}-only parallaxes, both the RR Lyrae and Cepheid zero points and the TRGB method
will be independently calibrated with at least one order of magnitude more calibrators,  
each having unprecedented precision.

\articlefigure[trim=0 140 0 20, width=0.9\textwidth]{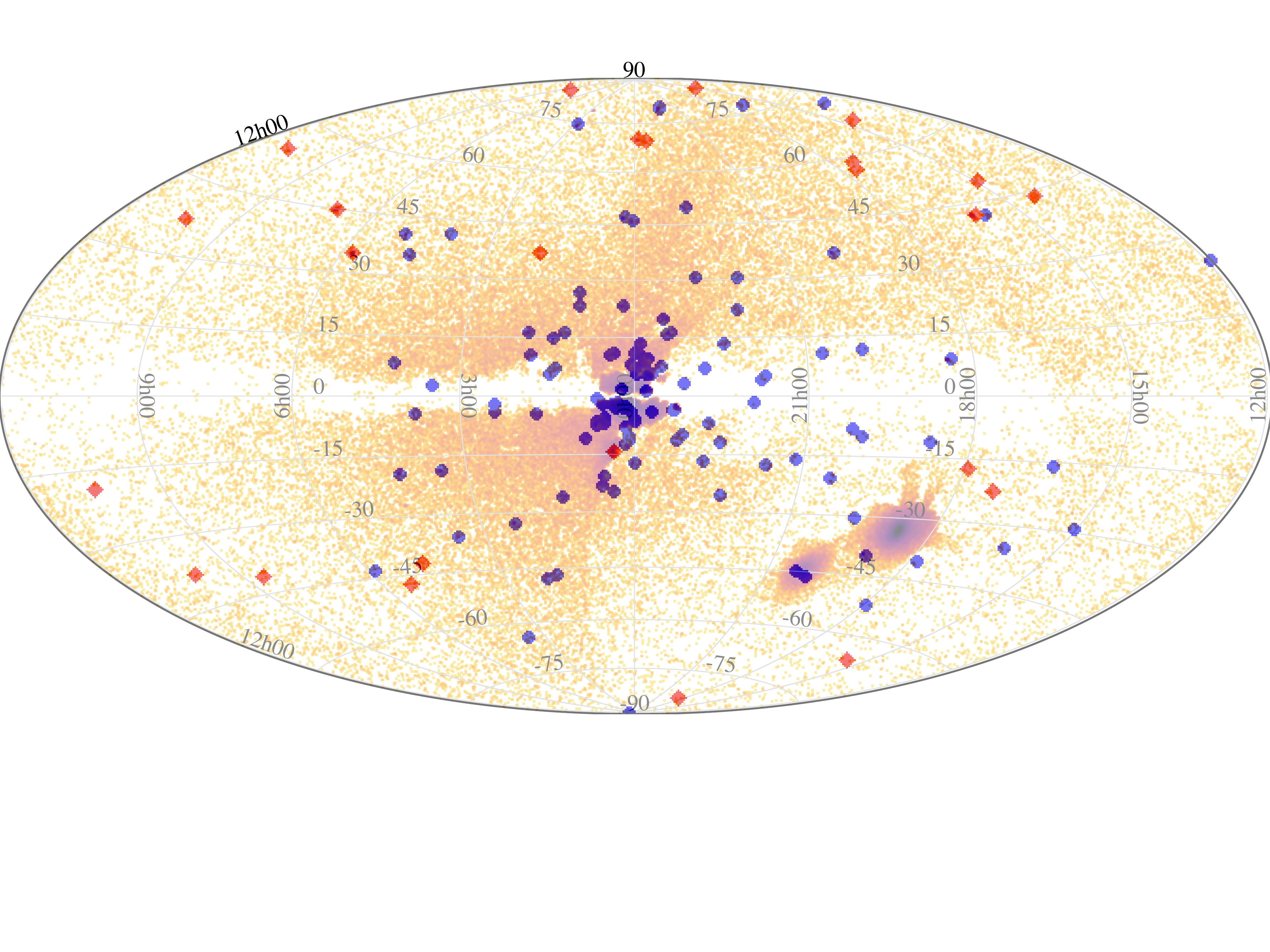}{fig4}{Distribution on the sky, in Galactic coordinates, of about 160,000 RR Lyrae stars known in the literature.  Filled blue dots and red filled diamonds indicate  globular clusters and dwarf spheroidal galaxies (classical and ultra-faint)  that are known to contain RR Lyrae stars.  See Fig.~45 of  \citet{clementini2018}
for an updated version of this figure showing new all-sky RR Lyrae stars released in  {\it Gaia} DR2 and literature variable stars, for a total of more than  223.000 sources.
}
\articlefigure[trim=0 340 0 40, width=0.9\textwidth]{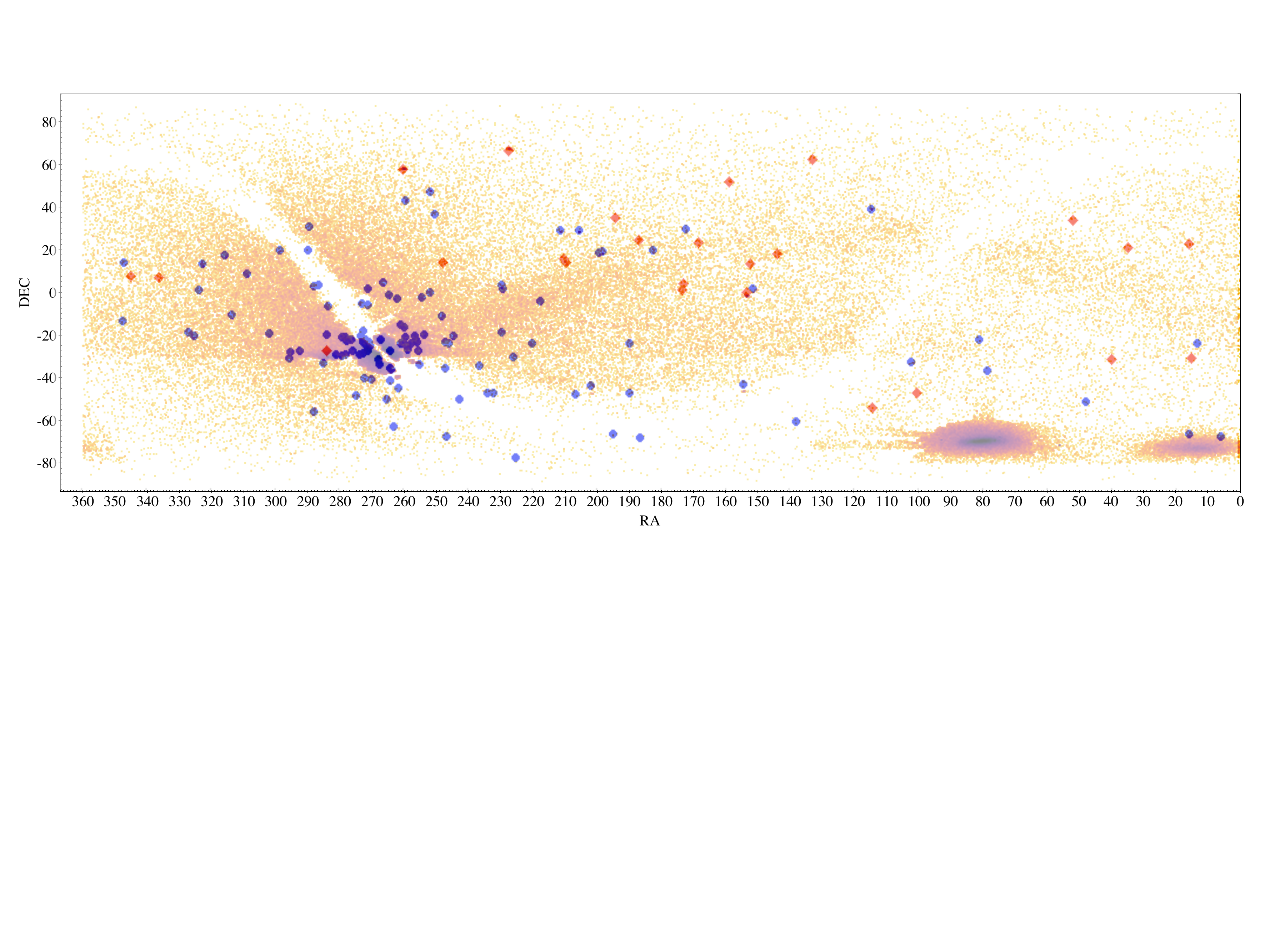}{fig5}{Same as in Fig.~\ref{fig4} but in equatorial coordinates.}

Figure~\ref{fig3} shows the distribution on the sky of known Cepheids.  More than 13,000 Cepheids of various types
 (Classical, Anomalous and Type~II) are shown in the figure. Over than 10,000  of these Cepheids were identified by the OGLE survey in the  Magellanic Clouds.
Figure~\ref{fig4}  shows a sky map, in galactic coordinates, of known RR Lyrae stars and, Fig.~\ref{fig5}, shows the same sample viewed in equatorial coordinates. 
About 160,000 fundamental-mode, first-overtone and double-mode RR Lyrae are shown, collecting data from many different surveys among which major ones   
are: OGLE, ASAS, PanSTARSS, CATALINA, LINEAR, ASASSN and {\it Gaia} DR1. Filled blue dots and red filled diamonds mark globular clusters and dwarf spheroidal galaxies (classical and ultra-faint)  that are known to contain RR Lyrae stars.
We look forward to see the fresh view of the MW halo, disk and bulge provided by the all-sky Cepheid and RR Lyrae maps released with {\it Gaia} DR2.

\section{Conclusions and Future Prospectives}
{\it Gaia}  second data release is going to deliver its promise of new and accurate astrometry (positions, parallaxes and proper motions) 
based only on  {\it Gaia} measurements  and of RR Lyrae stars and Cepheids identified across the whole Galaxy and its closest companions.
This will already allow a breakthrough in our understanding of the MW structure, dynamics, and formation, and a re-assessment of the cosmic distance ladder.
Results will further improve with {\it Gaia} DR3 in 2020 and with the release of {\it Gaia} final catalogue in 2022. 
Meanwhile, ESA has approved a 2-year extension of the {\it Gaia} mission, and further 
extensions of  2+1 years are likely to be approved in the future. This will leave  {\it Gaia} still fully operational  when  JWST will be launched, LSST will see its first light  (both events 
being currently  foreseen for 2020) and E-ELT first light will take place (currently foreseen for 2024), 
thus enhancing the synergy among  these outstanding projects.
Looking further ahead, a new Gaia-like mission is also being conceived with all-sky absolute optical and near-infrared (NIR) astrometry (GaiaNIR). 
Going into the NIR,   the new mission would be able to probe through the Galactic dust and most obscured regions of the MW disk, bulge and 
spiral arms.
 
In closing I would like to reflect on the great scientist to whom this meeting is dedicated. I met Jeremy more than twenty years ago, we shared the office in Bologna for about a  couple of months. We 
started a joint project to measure  the chemical  composition of bright Galactic RR Lyrae stars and study  their luminosity-metallicity relation. With that study we set the slope  of the $M_{V}$ - [Fe/H] relation, now {\it Gaia} will pin down its zero point with unprecedented precision. Thank you Jeremy, it has been an honour and a pleasure to work with you.

\acknowledgements 
This work makes use of data from the European Space Agency (ESA)
mission {\it Gaia} (\url{https://www.cosmos.esa.int/gaia}), processed by
the {\it Gaia} Data Processing and Analysis Consortium (DPAC,
\url{https://www.cosmos.esa.int/web/gaia/dpac/consortium}). 
Funding
for the DPAC has been provided by national institutions, in particular
the institutions participating in the {\it Gaia} Multilateral Agreement.
Support to this study has been provided by PRIN-INAF2014, "EXCALIBUR'S" (P.I. G. Clementini) 
and by Premiale 2015, "MITiC" (P.I. B. Garilli).


\begin{thebibliography}{}
\bibitem[Clementini et al.(2012)]{clementini2012} Clementini G., Cignoni, M., Contreras Ramos, R., et al.\ 2012, \apj,  756, 108
\bibitem[Clementini et al.(2016)]{clementini2016} Clementini G., Ripepi, V., Leccia, S., et al. 2016, \aap,  595, A133
\bibitem[Clementini et al.(2018)]{clementini2018} Clementini G., Ripepi, V., Molinaro, R., et al. 2018, \aap, submitted (arXiv: 1805.02079)
\bibitem[Delgado et al.(2018)]{delgado2018} Delgado H.~E., Sarro, L.~M., Clementini, G., Muraveva, T., Garofalo, A., \ 2018, \aap, submitted (arXiv: 1803.01162)
\bibitem[Eyer et al.(2017)]{eyer2017} Eyer, L., Mowlavi, N., Evans, D.~W., et al. \ 2017, \aap, submitted (arXiv:1702.03295)
\bibitem[Gaia Collaboration et al.(2016a)]{gai16a} Gaia Collaboration, Prusti, T., de Bruijne, J.~H.~J., et al.\ 2016, \aap, 595, A1 
\bibitem[Gaia Collaboration et al. (2016b)]{gai16b}Gaia Collaboration, Brown, A.~G.~A., Vallenari, A., et al. \ 2016, \aap, 595, A2
\bibitem[Gaia Collaboration et al.(2017)]{gai17} Gaia Collaboration, Clementini, G., Eyer, L., et al.\ 2017, \aap, 605, A79 
\bibitem[Lindegren et al.(2016)]{lin16} Lindegren, L., Lammers, U., Bastian, U., et al., 2016, A\&A, 595, A4 
\bibitem[Muraveva et al.(2015)]{muraveva2015} Muraveva, T.,  Palmer, M.,  Clementini, G., et al.\ 2015, \apj, 807, 127
\bibitem[Muraveva et al.(2018)]{muraveva2018} Muraveva, T., Delgado, H.~E., Clementini, G., et al.\ 2018, \mnras, submitted (arXiv:1805.08742)
\end{thebibliography}
\end{document}